\documentclass[aps,prl,times,twocolumn, footinbib, reprint, superscriptaddress, floatfix, longbibliography]
{revtex4-1}

\usepackage{amsmath,amssymb} % math symbols
\usepackage{bm} % bold math font
\usepackage{graphicx} % for figures
\usepackage{subcaption}
\usepackage{mathptmx}
\usepackage{float}
\usepackage{array}
\usepackage{comment} % allows block comments
\usepackage{textcomp} % This package is just to give the text quote '
\usepackage{enumitem}
\setlist{noitemsep,leftmargin=*,topsep=0pt,parsep=0pt}
\usepackage{xcolor} % \textcolor{red}{text} will be red for notes
\definecolor{lightgray}{gray}{0.6}
\definecolor{medgray}{gray}{0.4}
\usepackage{hyperref}
\hypersetup{
colorlinks=true,
urlcolor= blue,
citecolor=blue,
linkcolor= blue,
}
\newif\ifptitle
\newif\ifpnumber
\newcounter{para}

\ptitletrue  % comment this line to hide paragraph titles

\newcommand{\unit}[1]{\ensuremath{\, \mathrm{#1}}}
 \renewcommand{\vec}[1]{\bm{#1}}

\newcommand{\mytitle}{Machine learning magnetic interactions from neutron powder diffraction data}

\begin{document}

\title{\mytitle}

\author{Adit S. Desai}
\affiliation{School of Physics, Georgia Institute of Technology, Atlanta, Georgia 30332, USA}
\author{Yongqiang Cheng}
\email{chengy@ornl.gov}
\affiliation{Neutron Scattering Division, Oak Ridge National Laboratory, Oak Ridge, Tennessee 37831, USA}
\author{Joseph A.~M. Paddison}
\email{paddisonja@ornl.gov}
\affiliation{Neutron Scattering Division, Oak Ridge National Laboratory, Oak Ridge, Tennessee 37831, USA}
\affiliation{Materials Science and Technology Division, Oak Ridge National Laboratory, Oak Ridge, Tennessee 37831, USA}

\date{\today}

\begin{abstract}

Neutron diffraction is a versatile experimental technique capable of probing a material's magnetic properties. While diffraction is typically used to determine the magnetic structure of a material, magnetic diffuse scattering data from a diffraction experiment are also sensitive to the magnetic interactions in its Hamiltonian. However, accurately determining magnetic interaction parameters from neutron-scattering data involves an inverse scattering problem that is challenging to solve in general. Here, we investigate the effectiveness of a machine learning approach to predict the interaction parameters given magnetic diffuse-scattering data measured on powder samples, for a comprehensive survey of isotropic interactions on eight high-symmetry lattices. Across all lattices we considered, the machine-learning approach estimates the interaction parameters with high ($\sim$2\%) accuracy, while avoiding the issue of false minima that is encountered with non-linear least squares refinement. Our results highlight that powder diffuse-scattering data can provide a compact ``fingerprint" of the magnetic interactions for many materials.

\end{abstract}

\maketitle

\section{\label{sec:Start}Introduction}
Magnetism is responsible for behavior of both fundamental and applied significance. Examples include the soft magnets used in magnetic motors and transformers \cite{Silveyra_2018}, magnetocaloric materials for low-temperature magnetic cooling applications \cite{Gschneidner_2005}, and topological spin textures such as skyrmions that have potential applications for next-generation data storage devices \cite{Tokura_2020}. Moreover, magnetically frustrated systems \cite{ramirez_frustration} can give rise to exotic states of matter such as spin ices \cite{spin_ice_castelnovo, bramwell_spin_ice}, spin glasses \cite{spin_glass}, and spin liquids \cite{Balents_sl}. Of particular interest is the quantum spin liquid, which would not only provide valuable insight into fundamental physics but could also be applicable to future technologies such as quantum computing \cite{Kitaev_2003}.

\begin{figure*}[!htb]
    \centering
    \begin{subfigure}{0.215\textwidth}
        \includegraphics[width =\textwidth]{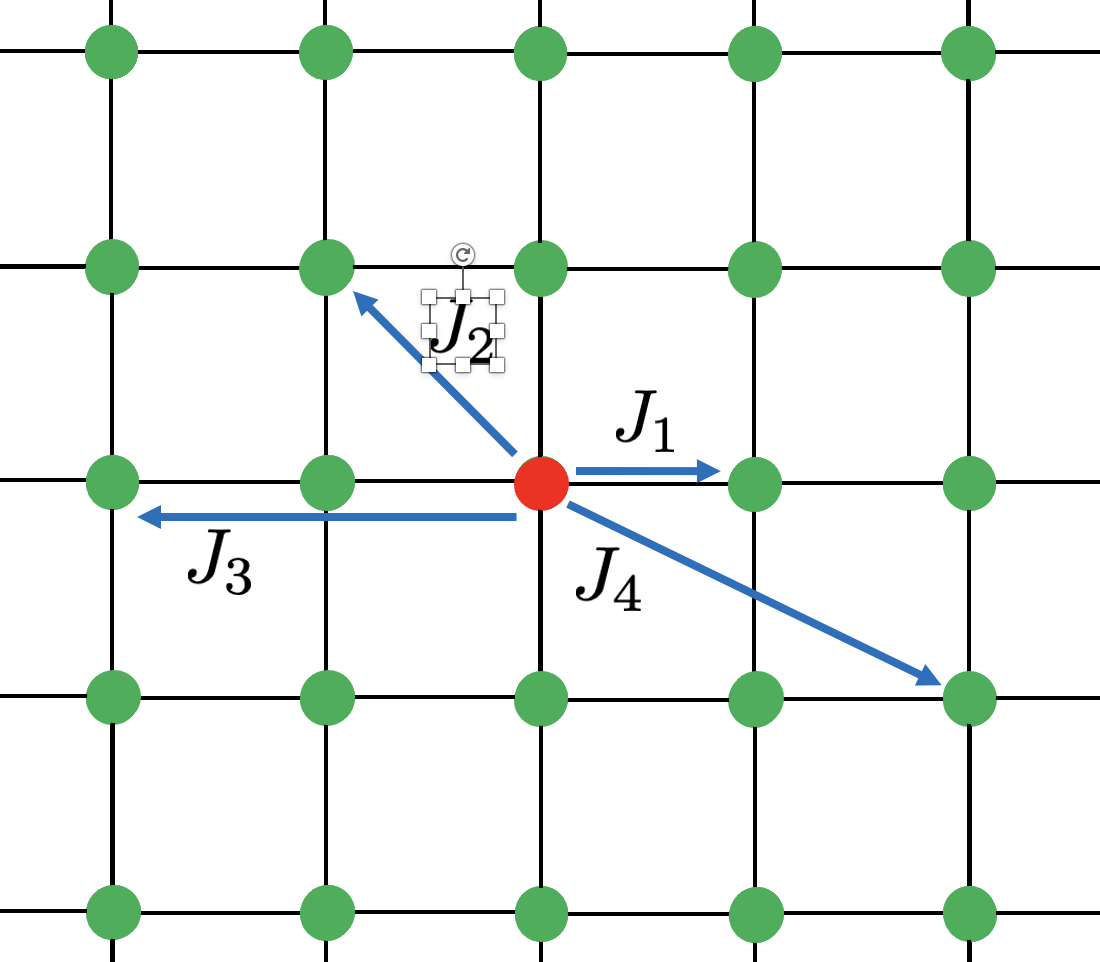}
        \subcaption{Square}
    \end{subfigure} 
    \begin{subfigure}{0.22\textwidth}
        \includegraphics[width = \textwidth]{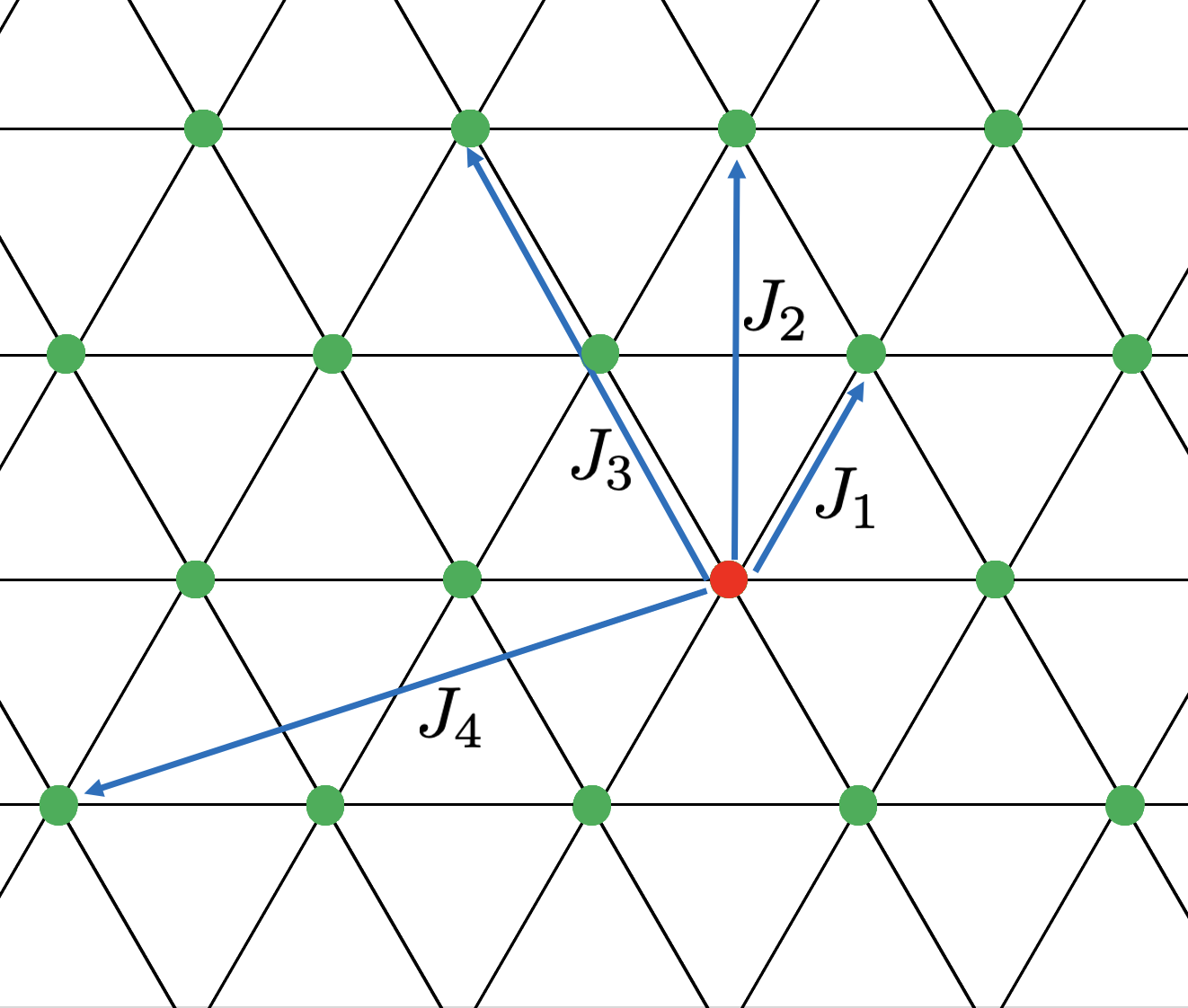}
        \subcaption{Triangular}
    \end{subfigure}
    \begin{subfigure}{0.22\textwidth}
        \includegraphics[width = \textwidth]{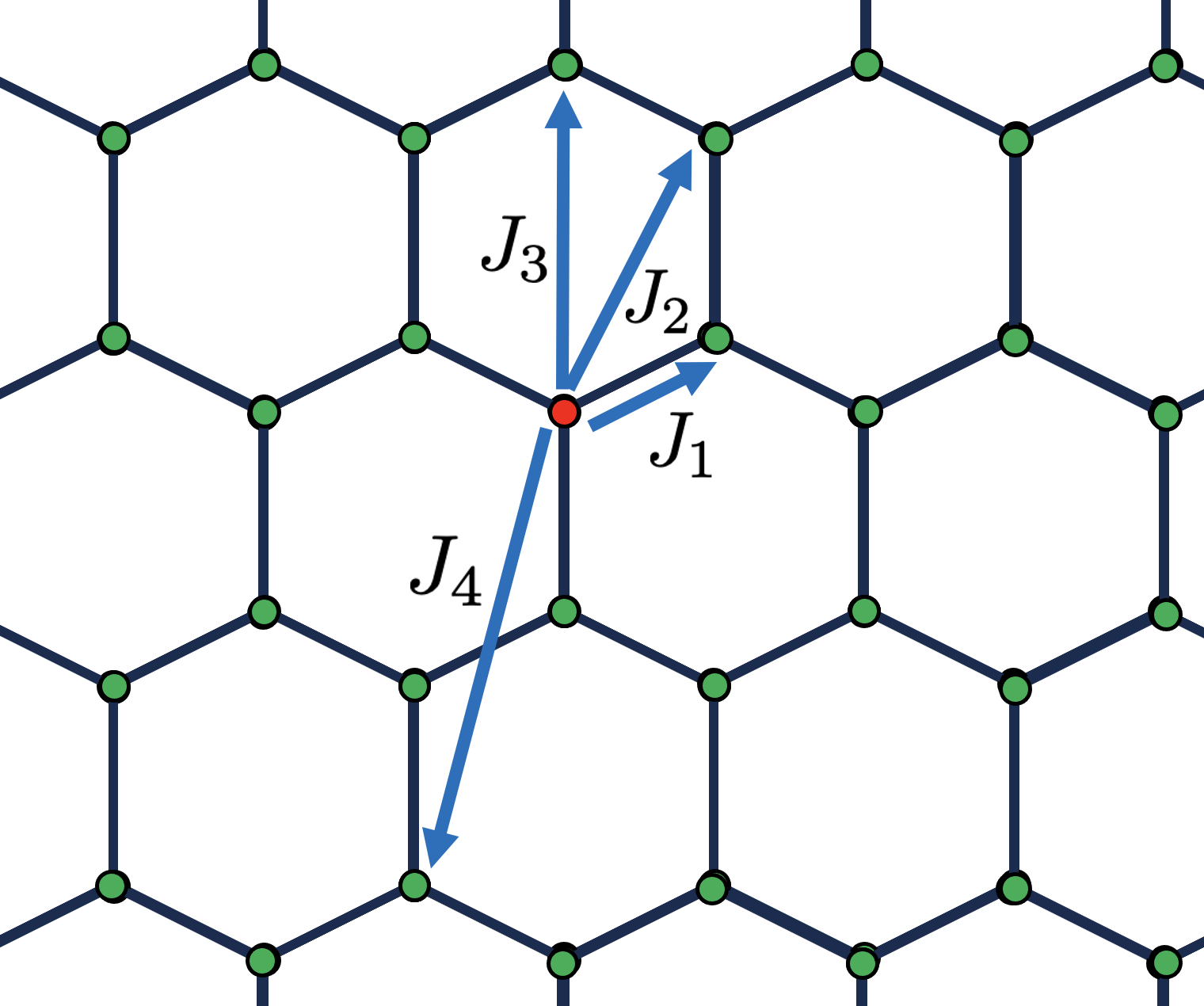}
        \subcaption{Honeycomb}
    \end{subfigure}
    \begin{subfigure}{0.187\textwidth}
        \includegraphics[width = \textwidth]{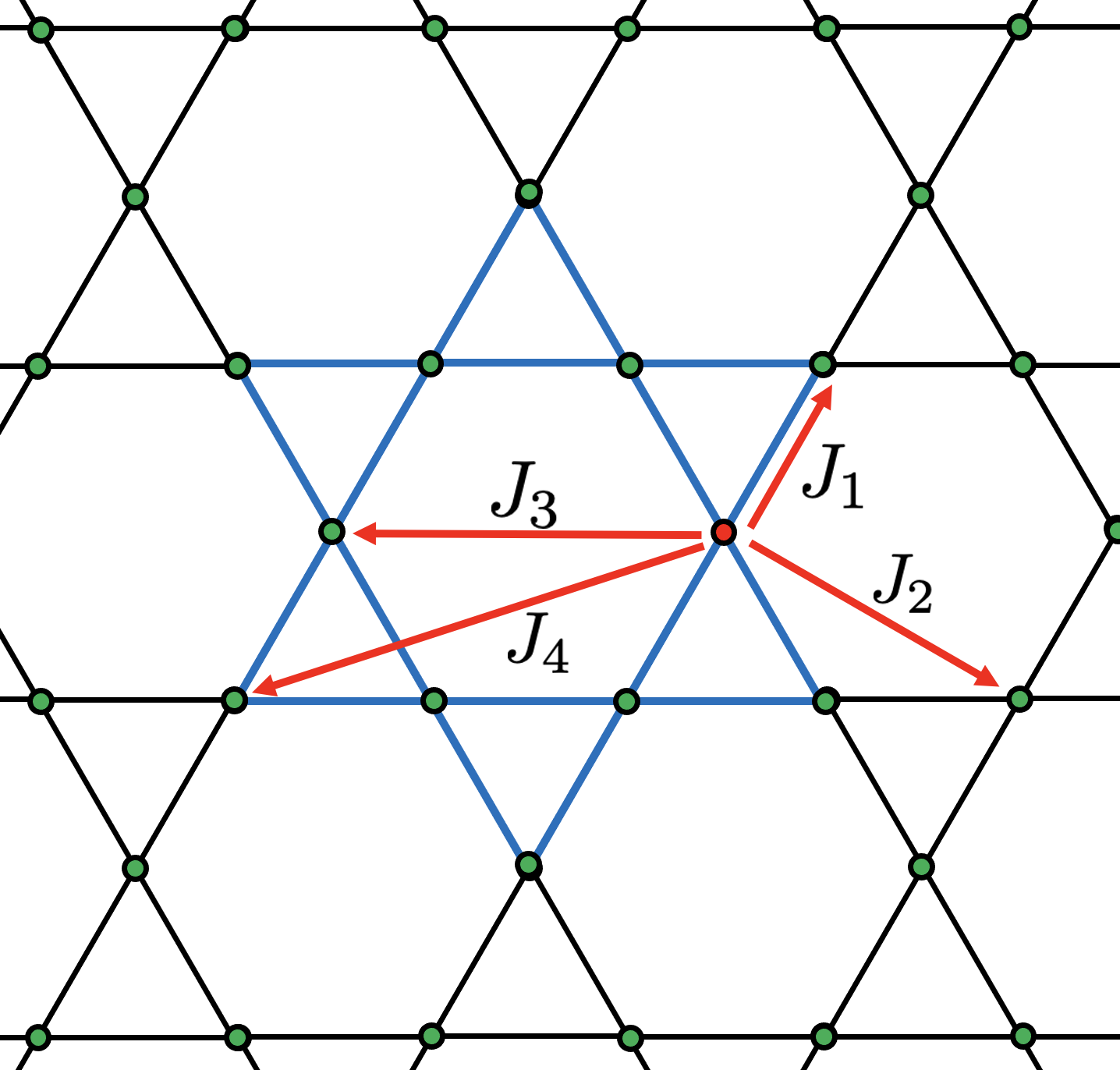}
        \subcaption{Kagome}
    \end{subfigure}
    \begin{subfigure}{0.22\textwidth}
        \includegraphics[width = \textwidth]{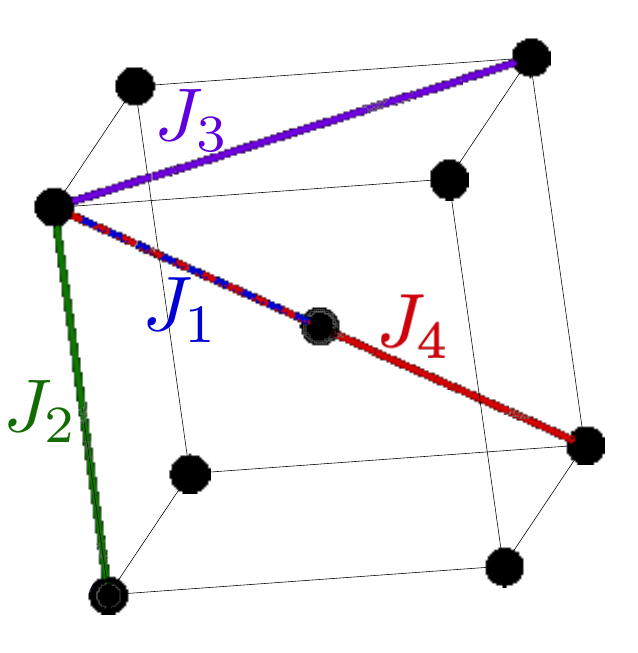}
        \subcaption{Body-Centered Cubic (BCC)}
    \end{subfigure}
    \begin{subfigure}{0.22\textwidth}
        \includegraphics[width = \textwidth]{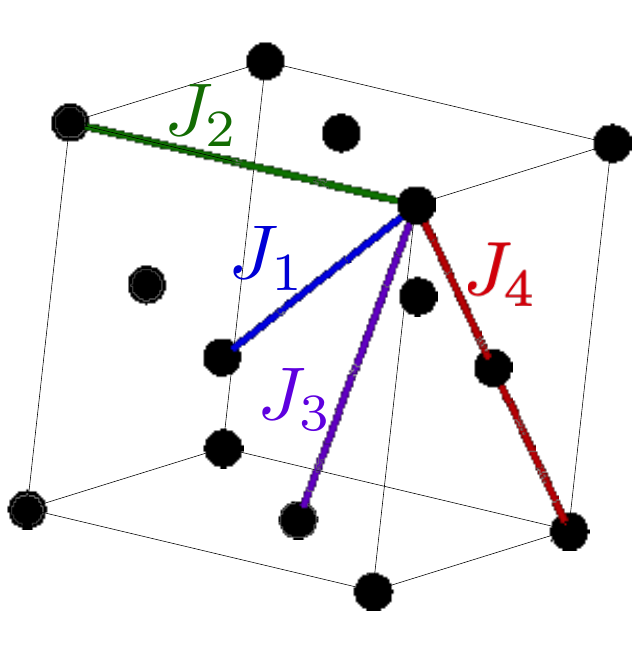}
        \subcaption{Face-Centered Cubic (FCC)}
    \end{subfigure}
    \begin{subfigure}{0.22\textwidth}
        \includegraphics[width = \textwidth]{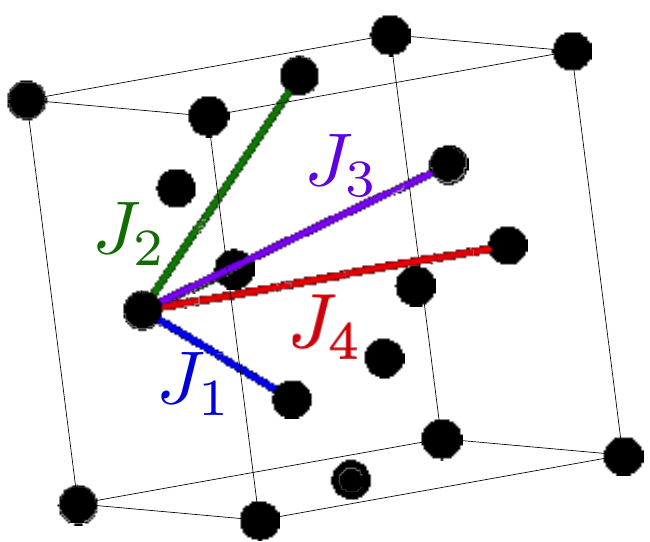}
        \subcaption{Diamond}
    \end{subfigure}
    \begin{subfigure}{0.22\textwidth}
        \includegraphics[width = \textwidth]{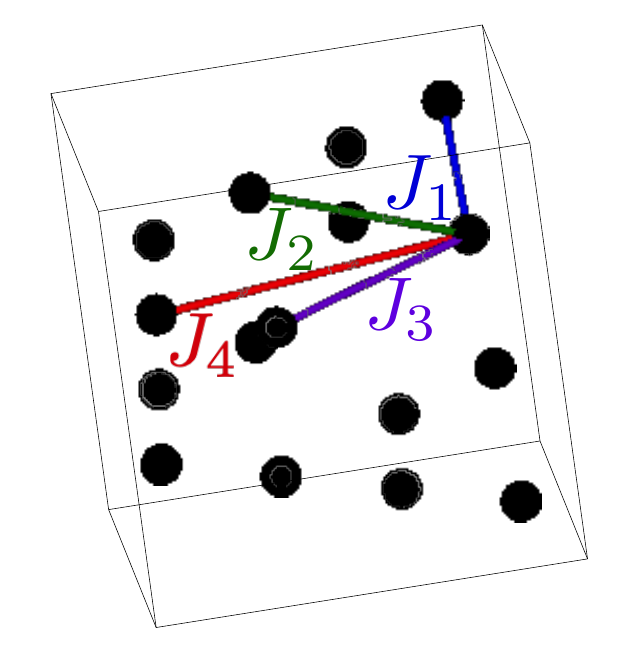}
        \subcaption{Pyrochlore}
    \end{subfigure}
    \caption{The lattices studied in this paper and their four near neighbor magnetic interactions with pathways labeled $J_1$, $J_2$, $J_3$ and $J_4$. Note that the kagome and pyrochlore lattices have two types of symmetry-inequivalent third neighbors that can have different interactions; one of these is labeled $J_3$ in the figure and the other is at double the vector corresponding to $J_1$. In this work we assumed that both third-neighbor interactions are the same.}
    \label{fig:Lattices}
\end{figure*}

To optimize a material's magnetic properties for technological applications, or to determine whether it hosts an exotic state of matter, it is often necessary to determine the magnetic interactions of that material. Neutron-scattering experiments are an excellent probe of magnetic materials, due to the neutron's intrinsic magnetic moment, ability to penetrate complex sample environments, and energy that can be tuned to match the energy scale of magnetic interactions \cite{Boothroyd}.
Extensive insights have been obtained from inelastic neutron-scattering data measured on single crystals, which, when measured below a material's magnetic ordering temperature, provide access to spin-wave dispersion curves that can determine the magnetic interaction parameters in its Hamiltonian \cite{Boothroyd,spinw,Fishman_2018}. 
An alternative method makes use of neutron diffraction data measured \textit{above} the material's magnetic ordering temperature \cite{blech1964spin,Lindgard_2007,paddison2022spinteract}. These data are sensitive to the magnetic interactions \emph{via} the modulation of the diffuse magnetic intensity with wavevector. This technique has been used to study spin-ice behavior \cite{Morris_2009}, spin-glass behavior \cite{Roth_2019}, and other complex magnetic states \cite{Lindgard_2007,Manuel_2009}, and has provided insight into the mechanisms of responses such as the topological Hall effect \cite{Paddison_2022}, the magnetocaloric effect \cite{Bulled_2022}, and colossal magneto-resistance \cite{Osborn_1998,Ye_2022}.
 
Irrespective of the scattering technique, inferring the magnetic interaction parameters from experimental data poses fundamental challenges that define an \emph{inverse scattering problem} \cite{Samarakoon_2022a}. The first challenge is that experimental data contain incomplete information about the underlying interactions; consequently, when it is possible to estimate the scattering data given a model Hamiltonian, the converse may not be true. The second challenge is that, even when the data are sufficiently information-rich, extracting this information effectively is not straightforward. Common approaches have involved qualitative comparisons of model calculations with experimental data, or nonlinear least-squares regression to estimate interaction parameters. The former approach relies strongly on the models selected for comparison, while in the latter case, a poor initial guess of the solution can cause a least squares regression to become trapped in a ``local minimum" far from the true solution \cite{Demidenko_1989}.

Recent advances in machine-learning algorithms provide a promising framework to address the inverse scattering problem \cite{Baldi_1989,Samarakoon_2022a}. Machine-learning methods have been employed for several important problems in magnetism, including the estimation of magnetic interactions from microscopy images \cite{Wang_2020,Kwon_2020} and prediction of magnetic ordering patterns \cite{Acosta_2022,Merker_2022}. 
Machine-learning algorithms have also been employed to estimate magnetic interactions from single-crystal inelastic and diffuse magnetic scattering data from spin-ice, manganite, and Kitaev honeycomb materials \cite{Samarakoon_2022,Butler_2021,Samarakoon_2020}.
In all cases, however, the approach was tailored to the system and experimental data of interest. It remains an open question whether machine learning methods can be employed to solve the inverse scattering problem for a wide range of magnetic interactions on different lattices, reflecting the diversity of magnetic states discovered experimentally.

Here, we explore how effectively machine-learning methods can solve the inverse scattering problem from magnetic diffuse-scattering data measured on powder samples. We consider an extensive survey of magnetic systems that includes 8 high-symmetry magnetic lattices and $\sim$$1.3\times 10^6$ sets
of interaction parameters per lattice, and train a neural network using simulated \emph{powder} diffuse-scattering data for each lattice. Our choice of powder diffuse-scattering data is motivated by the fact that they are compact and inexpensive to simulate compared to single-crystal inelastic scattering data. Despite the compact nature of powder data, we find that they are rich in information about the magnetic interactions. We further show that these interactions can usually be effectively extracted using either a neural-network method or by least-squares regression. However, the neural-network approach more effectively avoids false solutions and hence provides an attractive method to obtain initial estimates of the parameter values.

Our paper is structured as follows. We begin by describing the simulation of the training data, and assess the information content of these data.
We describe the construction of the neural network, and the training process. We compare the performance of the neural network to conventional nonlinear least-squares regression. Finally, we report the results of a test of neural network on experimental data from the compound CoRh$_2$O$_4$ \cite{experimental_data}.

\section{\label{sec:Methods} Methods}

The first step in developing a neural network is to build an extensive set of training data. Unfortunately, the quantity of experimental data available for materials with accepted magnetic interaction values is significantly lower than what is needed. Additionally, training a neural network on experimental data could make it vulnerable to experimental noise. We therefore trained the neural network on simulated powder data built using the program Spinteract \cite{paddison2022spinteract}, which uses a field-theoretic approximation \cite{Brout_1967,Logan_1995,Conlon_2010} to simulate diffuse scattering data. This method is computationally inexpensive, typically requiring less than 1\,s to simulate a data set, but has accuracy comparable to Monte Carlo simulations at elevated temperatures relevant for magnetic diffuse-scattering measurements \cite{Conlon_2010}.

In this paper, we exclusively consider the isotropic Heisenberg model, in which the energy of the magnetic system is given by \cite{ramirez_frustration}
\begin{equation}
    H = -\frac{1}{2} \sum_{n} J_n \sum_{i, j\in n} \vec{S}_i \cdot \vec{S}_j,
\end{equation}
where $J_n$ is the interaction parameter coupling spins $\mathbf{S}_i$ and $\mathbf{S}_j$, which are separated by an $n$-th neighbor distance. The spins $\mathbf{S}_i$ are modeled as classical vectors of unit length. Note that in this formalism ferromagnetic interactions correspond to positive $J_n$ values. Though this is a simplified interaction model, which neglects anisotropic interactions, it provides an effective description of many materials where the magnetism arises from transition-metal ions \cite{Rosner_2002,Hohlwein_2003,van1995heisenberg_example_3, Shirata_2012, Babkevich_2016}.

\begin{figure}
    \centering
    \includegraphics{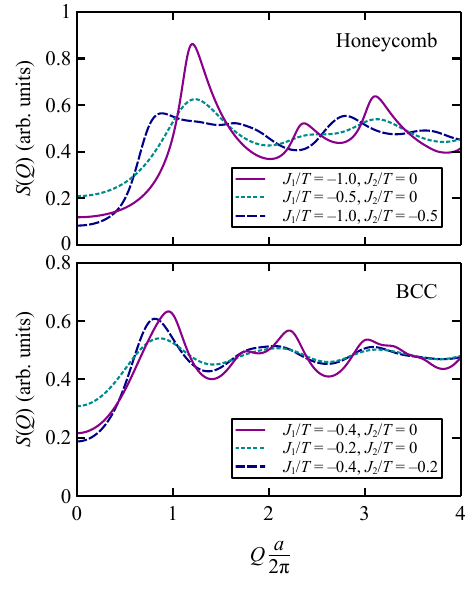}
    \caption{Simulated magnetic diffuse $S(Q)$ for the honeycomb lattice (top panel) and BCC lattice (bottom panel), with values of the interaction parameters labeled on the plots.} %

    \label{fig:example_Curves}
\end{figure}

We considered 8 high-symmetry crystal lattices known to resemble interesting materials (see, e.g., \cite{Chamorro_2018, Paddison_2016, Yi_2008, boulet2021_bcc, Karunadasa_2003, Han_2012, Yamura_2012, Sears_2015}). For each lattice, we considered four magnetic interactions that couple spins up to fourth nearest neighbors ($n=4$). The studied crystal structures and interaction pathways are shown in Figure~\ref{fig:Lattices}.
Diffuse-scattering data $S(Q)$ were simulated as a function of the dimensionless variable 
\begin{equation}
Q^{\prime} = \frac{aQ}{2 \pi},
\label{eq:q_prime}
\end{equation}
where $a$ is the lattice parameter and $Q$ is the magnitude of the wavevector. These simulated data are relevant to any material with the same lattice, irrespective of its lattice parameter. We note, however, that this approach only applies to high-symmetry lattices such as those shown in Figure~\ref{fig:Lattices}.

Figure \ref{fig:example_Curves} shows representative examples of the simulated diffuse $S(Q)$ for different lattices and interactions. We highlight some key points about these data. First, diffuse scattering contains broad features rather than sharp peaks. This is because diffuse scattering is obtained above the system's magnetic ordering temperature, where only local magnetic correlations exist. Second, despite the diffuse nature of the scattering, differences are clearly apparent between different lattices and interactions. Specifically, the \emph{positions} of the diffuse maxima are determined by the choice of lattice and interactions. Third, the \emph{width} of the diffuse maxima is determined by the temperature of the simulation, relative to the energy of the magnetic interactions expressed in temperature units. 
We therefore define the dimensionless \emph{reduced interaction} as
\begin{equation}
\tilde{J}_{n} = \frac{J_n}{T}.
\label{eq:j_tilde}
\end{equation}
\begin{figure*}
    \centering
    \includegraphics[width = \textwidth]{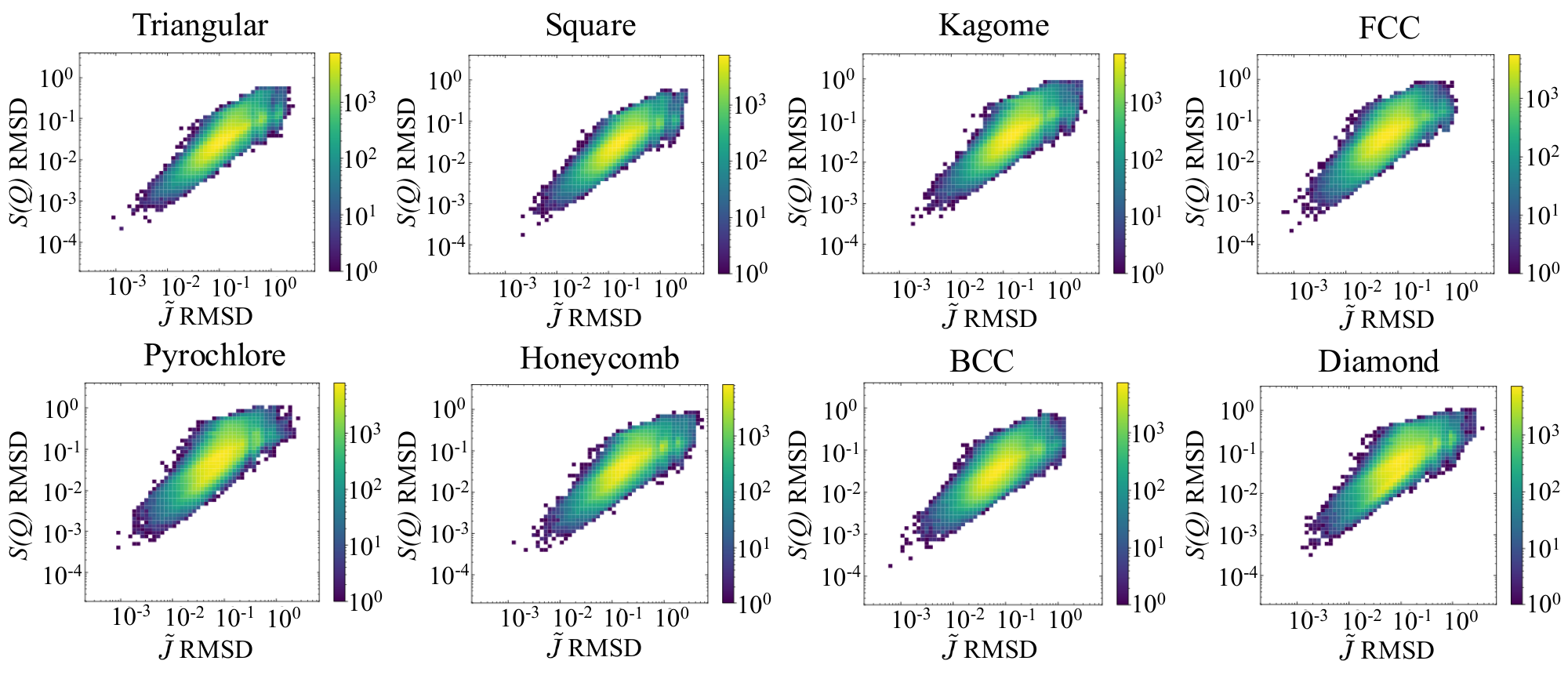}

    \caption{Graphs demonstrating sensitivity of the simulated diffuse intensity $S(Q)$ to differences in the magnetic interactions $J_n$. In each panel, the root-mean-square deviation (RMSD) in $S(Q)$ for a given lattice is plotted against the RMSD in the interaction values $\tilde{J}_n$, for randomly-chosen pairs of simulations in the training data. Probability density is shown in false color.}
    \label{fig:sq_v_parameter}
\end{figure*}

We simulated data according to the following method, which was motivated by the need to provide an extensive survey of the interaction space with a tractable  number of simulations.
First, we chose the nearest-neighbor interaction $J_1$ to be either $-1$ or 1\,$\unit{ K}$, and the further neighbor interactions $J_2$, $J_3$, and $J_4$ were initially chosen on a grid with constant steps: $J_2$  was simulated from $-3$ to $3$\,K in steps of $0.2$\,K, $J_3$ was simulated from $-2$ to $2$\,K in steps of $0.1$\,K, and $J_4$ was simulated from $-1$ to $1$\,K in steps of $0.1$\,K. We then used an adaptive model to add more training data near sensitive locations. If there was a large difference in $S(Q)$ between two neighboring sets of interaction parameters, then an additional simulation with parameters midway between the two interaction values was added to the training library.
For each set of interaction parameters $\{ J_n \}$, $S(Q)$ was simulated at 16 temperatures with fixed ratios of the average interaction magnitude, which we define by analogy with the Curie-Weiss temperature \cite{paddison2022spinteract} as
\begin{equation}
    T^{\ast} = \frac{1}{3} \sum_n Z_n|J_n| ,
\end{equation}
where $Z_n$ is the coordination number for the $n$-th neighbor shell. 
Finally, for each set of interactions $\{ J_n \}$ and temperature $T$, the corresponding set of reduced interactions $\{ \tilde{J}_n \}$ is obtained from Eq.~\eqref{eq:j_tilde}.
Simulations at temperatures below the ordering temperature were discarded. This resulted in an average of $\sim 1.3 \times 10^6$ simulations per lattice, with a minimum of $1.1 \times10^6$ simulations for the face-centered cubic lattice and a maximum of $1.6 \times10^6$ simulations for the diamond lattice. 

\section{\label{sec:Results} Results}

Before building the neural network, it is important to verify whether machine learning would be useful as an analysis approach. The question is whether diffuse-scattering data measured on powder samples are sufficiently information-rich to allow the inverse scattering problem to be solved. The answer is not yet known in general, since the method has only been applied to specific materials. 

To answer this question, we shuffled all the simulations in the training data for a given crystal structure and formed random pairs. For each pair of these simulations ($\sim 5\times 10^5$ pairs in total), we calculated the root-mean-square difference (RMSD) of $S(Q)$
and the RMSD of $\tilde{J}_n$. These quantities are defined as
\begin{equation}
\textrm{RMSD}(\tilde{J_n}) = \frac{1}{2}{\sqrt{\sum_i^4 (\tilde{J}_i^{(1)} - \tilde{J}_i^{(2)})^2}},
\end{equation}
\begin{equation}
\textrm{RMSD}[S(Q)] = {\sqrt{\frac{1}{N_Q}\sum_i^{N_Q} ({S}_i^{(1)}-{S}_i^{(2)})^2}},
\end{equation}
where $N_Q =79$ is the number of $Q'$ points between 0.1 and 4.
Figure~\ref{fig:sq_v_parameter} shows a correlation plot that demonstrates the magnitude and direction of the relationship between RMSD($\tilde{J}_n$) and RMSD[$S(Q)$]. For all lattices, there is a clear trend that larger parameter differences result in larger differences in $S(Q)$, and there is no appreciable probability of a large change in interactions producing a negligible change in $S(Q)$. These results are encouraging, as they indicate that $S(Q)$ data are highly sensitive to changes in the interactions. However, the precise relationship is nontrivial and has a large variance, which implies that the machine learning method is likely to be beneficial.
We emphasize that these promising results would not be obtained for other plausible choices of data. For example, if the ordered spin arrangement at $T=0$ was used instead of diffuse $S(Q)$ data, the correlation plot would reveal that the ordered spin arrangement can remain identical when the magnetic interactions are changed; this occurs unless a phase boundary is crossed.

\begin{figure*}
    \centering
    \includegraphics[width = 0.9\textwidth]{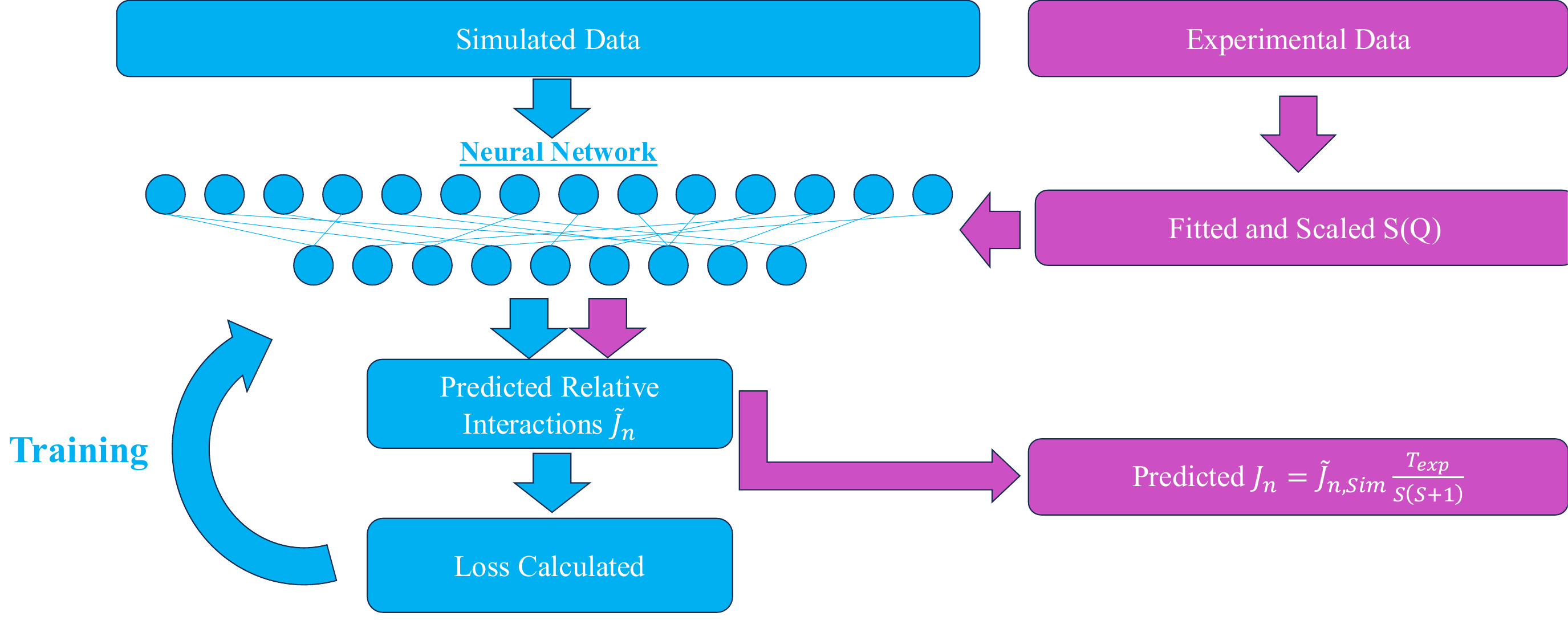}
    \caption{Schematic for training the neural network and applying it to experimental data. Blue blocks on the left show the process of training the neural network, and purple blocks on the right show the application to experimental data, which must first be fitted to a smoothing function and normalized before being input to the network.}
    \label{fig:flowchart}
\end{figure*}

\begin{figure*}
    \centering
    \includegraphics[width=0.9\textwidth]{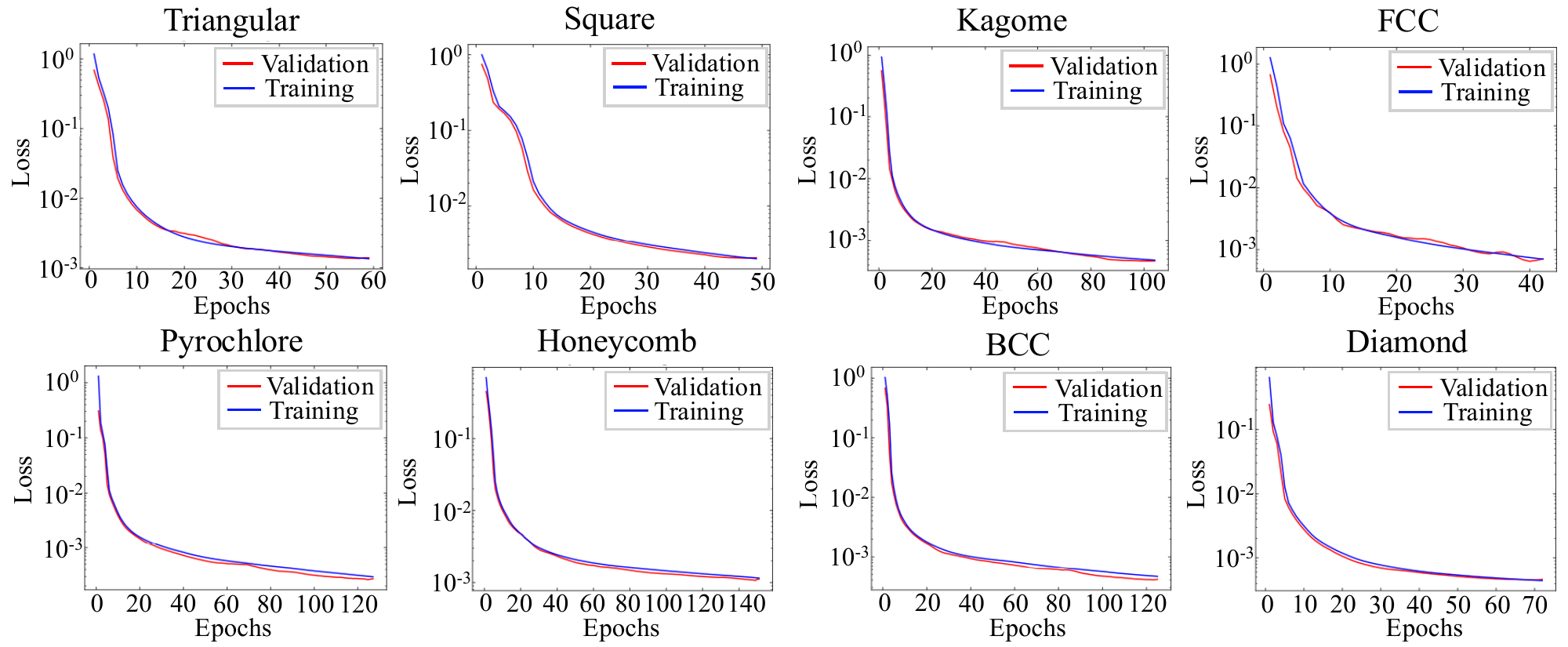}
    \caption{Learning curves of the neural network for validation data (red lines) and training data (blue lines). Each panel shows the loss defined in Eq.~\eqref{eq:loss} as a function of the number of completed training loops (epochs) for the lattice indicated. }
        \label{fig:diamond learning}
\end{figure*}

After conducting this test, a multilayer perceptron (MLP) neural network was built from Pytorch \cite{paszke2019pytorch}. The network architecture used three linear transform layers which sandwich rectified Linear Units layers (ReLU) as defined from Pytorch's library. While more sophisticated approaches, such as a convolutional neural network, were attempted, we found no significant improvement in performance.  
Before training the neural network, the data for a given crystal structure and temperature were normalized such that the median $S(Q)$ for all simulations was 1. 
The data were randomly shuffled and split into training, validation, and testing data. We used 80\% of the data as training data, 10\% were used as validation data, and the remaining 10\% were used as testing data. The training data are the data from which the neural network builds its model. The validation data are used to observe how the network performs while it is learning. The network will not learn from the validation data. Finally, the testing data are used to measure the performance of the network after it has finished training. 

The neural network was built using the Adam optimizer \cite{kingma2015adam} and trained using a modified version of the mean-squared-error (MSE) loss function \cite{sge_mse}, given by
\begin{equation}
   \text{Loss} =  \frac{1}{4} \sum^4_{i=1} \frac{(\tilde{J}_i^{\textrm{true}} - \tilde{J}_i^{\textrm{pred}})^2}{(\tilde{J}_1^{\textrm{true}})^2}.
   \label{eq:loss}
\end{equation}
The reason Eq.~\eqref{eq:loss} includes the $(\tilde{J}^{\textrm{true}}_{1})^2$ factor in the denominator is to ensure that the neural network is sensitive to simulations with smaller reduced interaction strengths.
Its architecture and hyperparameters were tuned to optimize the performance.

\begin{figure*}
    \centering
    \includegraphics[width = 0.9\textwidth]{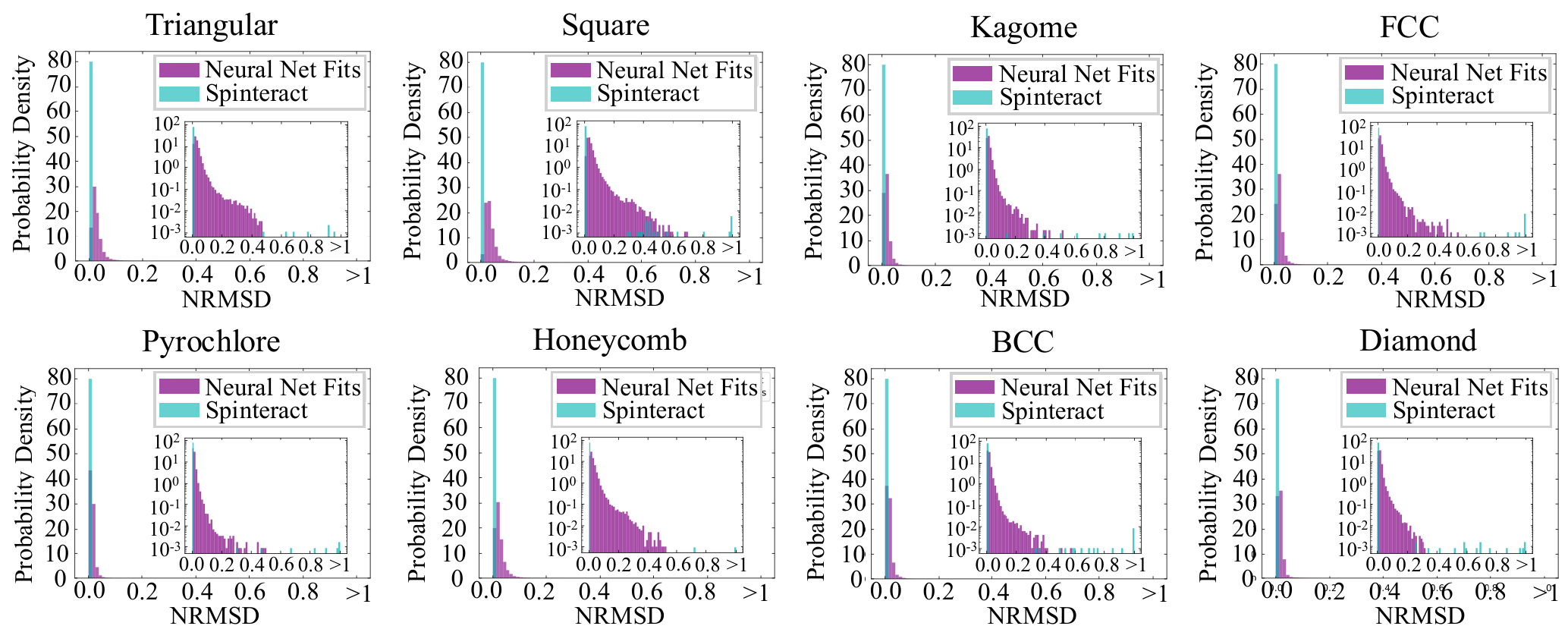}

    \caption{Performance of the neural network (purple bars) compared with nonlinear least-squares regression using the Spinteract program \cite{paddison2022spinteract} (cyan bars). For each lattice, performance is shown by plotting the probability-density function of differences between predicted and true values of the interactions obtained for the test data. The normalized RMS difference (NRMSD) is defined by Eq. \eqref{eq:NRMSD}. The average NRMSD for the least-squares regression is lower than for the neural network, but least-squares regression yield outliers with large NRMSD. Insets show the probability density on a log scale. Note that, to avoid an overly large tail, error values larger than 1 are shown in the bin labeled ``$>1$". }
    \label{fig:all lattices}
\end{figure*}

An outline of the neural network workflow is shown in Figure \ref{fig:flowchart}. To ensure the network was not overtrained, we studied how the network's training data and validation loss decreased with training loops, known as a learning curve. A key indicator of overtraining is a validation learning curve that stagnates or rises again. In order to prevent overtraining, the neural network would continue training until the validation loss would start to increase. We found this approach sufficient; while in some models the validation loss may start to decrease again if the model is trained further, we did not observe such behavior. The learning curves for all crystal structures are shown in Figure \ref{fig:diamond learning}. They reveal a good fit where the loss decreases for both training and validation data, and stabilizes at similar values for both.

The performance of the neural network on the test data is shown in Figure~\ref{fig:all lattices}. For a given lattice and set of interaction parameters, the error in the predicted interaction parameters is quantified by a modified version of root mean square difference, 
\begin{equation}
\textrm{NRMSD} = \frac{\sqrt{\sum_i^4 (\tilde{J}_i^{\textrm{true}} - \tilde{J}_i^{\textrm{pred}})^2}}{2 |\tilde{J}_1^{\textrm{true}}|}.
\label{eq:NRMSD}
\end{equation}
The mean value of this error, averaged across all lattices and sets of interactions, is 0.022.
This statistic demonstrates that machine learning is highly successful at estimating magnetic interaction parameters from magnetic diffuse scattering data. 

It is important to compare the results of the neural-network approach with those from traditional least-squares fitting. To do so, we used the same test data as input for nonlinear least-squares regression, as implemented in the Minuit program \cite{James_1975,James_1994}, with the four interactions $\{J_n \}$ as variable parameters. Initial values of all interactions were set to zero, as may occur if the user lacks prior knowledge of their expected values. An overall intensity scale factor was also refined. 
The performance of the least-squares and neural-network fits is compared in Figure~\ref{fig:all lattices}. Notably, the mean value of the error for the least-squares fits is $1.3\times10^{-4}$, which is significantly smaller than the corresponding value of 0.022 for the neural-net fits. Inspection of the distribution of errors reveals that the least-squares fits usually obtain precisely correct values, whereas the neural network approach has a distribution of values centered close to the correct answer [Figure~\ref{fig:all lattices}]. The tails of the distribution reveal another trend (see insets of Figure~\ref{fig:all lattices}): The least-squares regression occasionally yields outlying parameter values that are far from the correct answer. These outliers correspond to the least-squares regression failing to converge to the true minimum and are not observed when using the neural net. This result suggests that the a particularly useful application of the neural-network approach is to provide an initial estimate of the interaction values if prior knowledge of them is limited. This initial estimate may be subsequently improved by least-squares regression.

Experimental data on real materials contain statistical noise, as well as systematic errors arising from non-magnetic scattering. 
Therefore, it is crucial to assess how effectively the neural-net approach can solve the inverse-scattering problem from experimental data. We consider the compound $\unit{CoRh_2O_4}$, in which magnetic Co$^{2+}$ ions with spin $S=3/2$ occupy a diamond lattice and undergo long-range magnetic ordering at a temperature of $25$\,K \cite{experimental_data}. This material was chosen because its Hamiltonian is well understood from a previous inelastic neutron-scattering study, which concluded that the antiferromagnetic nearest-neighbor interaction  $J_1 = -0.63$\,meV and all other interactions are negligible \cite{experimental_data}. Experimental diffuse-scattering data were extracted from published inelastic neutron-scattering data \cite{experimental_data} measured at $T_\textrm{exp}=40$\,K by integrating the measured intensity over energy transfer between 0.4 and 6.0\,meV. 

Before providing the experimental data to the neural network, the data first had to be treated so that the network could interpret them correctly. First, the experimental data were divided by the square of the magnetic form factor of Co$^{2+}$, which was obtained from tabulated values \cite{Brown_2004}. 
Second, it is necessary to smooth the data to prevent the neural network from overfitting to statistical noise. This was achieved by fitting the intensity modulation to the function
\begin{equation}
    S_{\textrm{smooth}}(Q) = a_0 + \sum_{n=1}^4 a_n \frac{\sin(Q r_n)}{Q r_n},
\label{eq:fitting}
\end{equation}
where $r_n$ is the distance between $n$-th nearest neighbors and $a_n$ and $a_0$ fitting parameters \cite{blech1964spin,Park_2003}.
Third, we normalized the intensity values to their median, as was done for the simulated data. Finally, we converted $Q$ to $Q^{\prime}$ using Eq.~\eqref{eq:q_prime} and the lattice constant $a=8.50 \unit{\AA}$ \cite{experimental_data}, and resampled the fitted experimental data to the same $Q^{\prime}$ grid as the training data.

The predicted interaction values were converted into measurable interaction values by applying the relation
\begin{equation}
    J_n = \tilde{J}_{n, \mathrm{pred}} \frac{T_{\mathrm{exp}}}{S(S+1)}.
\end{equation}
The factor of $S(S+1)$ occurs because experimental values of $J$ were quoted for pairwise interactions between classical spins of length $\sqrt{S(S+1)}$ \cite{experimental_data}, whereas our training data assume spins of unit length. The network predicted $J_1=-0.555$, $J_2=-0.077$, $J_3=-0.043$, and $J_4=-0.039$\,meV.This prediction is close to the results of Ref.~\cite{experimental_data}, and correctly identifies $J_1$ as the dominant interaction. 
After inputting the neural-network solution to a least-squares regression, the final predicted $J_1=-0.63(2)$, $J_2=-0.09(1)$, $J_3=-0.05(1)$, and $J_4=-0.03(3)$ meV. This agrees very well with the previous experimental results obtained from analysis of spin-wave spectra \cite{experimental_data}.  

\begin{figure}
    \centering
    \includegraphics[width = 0.45\textwidth]{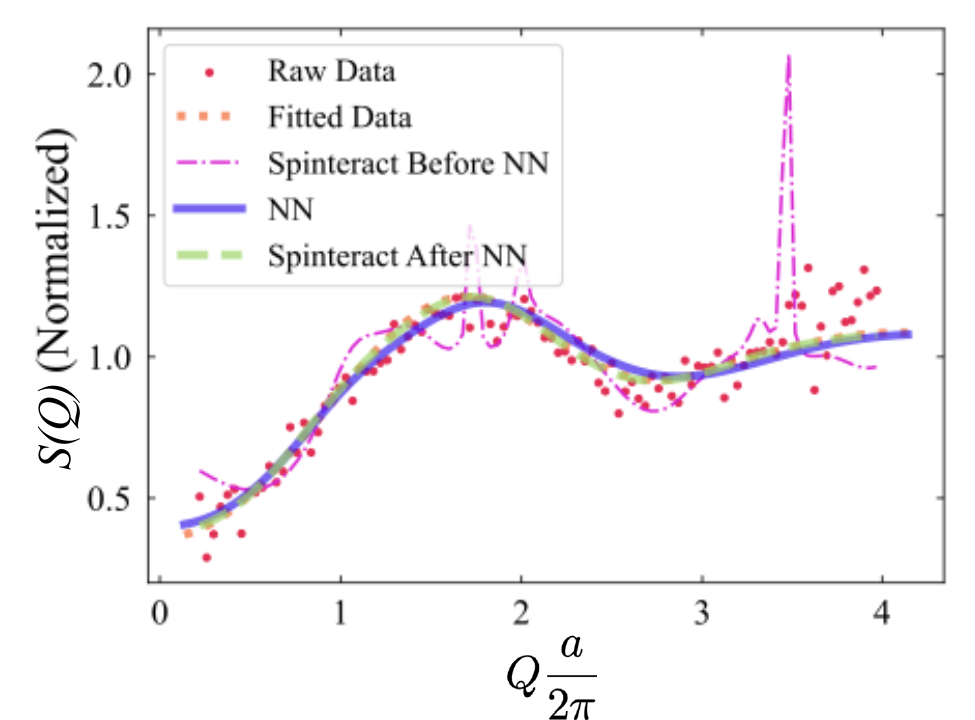}
    \caption{
   Stages of testing the neural network on experimental magnetic diffuse-scattering data measured on the compound \unit{CoRh_2O_4} at a temperature of 40\,K. First, the raw data (red points) were fitted to Eq.~\eqref{eq:fitting} to obtain the smoothed fitted data (orange dashed line), which were input to the neural network to obtain the predicted curve (blue solid line); the predicted $J_1 = -0.55$ meV, with further interaction magnitudes $<0.08$\,meV. This prediction was then input as an initial guess for nonlinear least-squares regression to obtain the optimized fit (green dashed line); the optimized $J_1 = -0.63$\,meV, with further interaction magnitudes $<0.09$\,meV. For comparison, nonlinear least-squares regression with a poor initial guess of the interactions yields an unsatisfactory fit (pink dot-dashed line) with unphysical interaction values (see text).}
     \label{fig:CoRh2O4 New}
\end{figure}

To illustrate the utility of having an initial guess provided by the neural network, we also attempted a least-squares regression from a poor initial guess of the interactions. Having started from an initial guess of $J_1 = -1$, $J_2 = 1.8$, $J_3 = 0.36$, and $J_4 = -0.72$ meV, the least-squares refinement became trapped in a local minimum, leading to a very poor prediction of $J_1 = -54.56$, $J_2 = -10.14$, $J_3 = -2.91$, and $J_4 = 47.44$ meV. 
The data, the smoothed data input into the neural network, the neural network's initial prediction, the least-squares refinement using the neural network's guess, and the least-squares refinement without using the neural network's guess, are compared in Figure~\ref{fig:CoRh2O4 New}.

\section{\label{sec:Figures} Conclusion}

Our study reveals three key results. First, magnetic diffuse scattering data measured on powder samples are rich in information about isotropic magnetic interactions. Based on an extensive survey of $\sim$$10^{7}$ total simulations and 8 different lattices, we find that changing the magnetic interactions consistently causes a change in the diffuse scattering. This change is generally sufficient to estimate isotropic magnetic interactions from powder data, enabling the inverse scattering problem to be solved in the isotropic case.

Second, we find that both neural-network fitting and least-squares regression are effective for extracting the magnetic interactions. 
Least-squares regression obtains accurate and precise results in most cases, but highly inaccurate results in a small minority of cases, whereas the neural-network approach produces less precise values but suffers fewer serious failures. This result suggests that the machine-learning approach may be most useful in providing starting values for least-squares regression.

Third, we have shown that this methodology can be applied to experimental data, and the interaction parameters obtained are in good agreement with the ``gold standard" approach of inelastic neutron scattering. Since powder neutron-diffraction experiments are often faster and more easily automated than single-crystal inelastic scattering measurements \cite{Losko_2014}, we anticipate that the former approach will support an increase in throughput of both collection and analysis of suitable data, assisting development of magnetic materials needed for future technology.

A limitation of our analysis is that we considered only isotropic interactions. While this approximation is often reasonable for magnetic compounds based on transition metals, it usually does not hold for rare-earth magnets, due to the effects of crystal field and spin-orbit coupling \cite{Skomski_2011}. Powder averaging reduces, but does not always remove, the sensitivity of diffuse-scattering data to anisotropic interactions \cite{Paddison_2020}; therefore, further work considering anisotropic interactions would be valuable.

Most importantly, our results suggest that powder diffuse-scattering data provide a ``fingerprint" of  isotropic magnetic interactions. Remarkably, this holds despite the compact nature of powder data. 
Our results also hint that machine-learning methods may hold the greatest promise for analyzing experimental data from multiple experimental probes simultaneously \cite{Doucet_2021,Chen_2021}. In particular, our analysis points towards the ability of these methods to converge to the vicinity of the correct solution while avoiding false minima, which is likely to become very important for multi-modal and multi-dimensional data sets.

\vspace{\baselineskip}
\vspace{\baselineskip}

\begin{acknowledgments}
We are grateful to Martin Mourigal for valuable discussions and for providing computational resources. We would like to thank the U.S. Department of Energy, Office of Science, Office of Workforce Development for Teachers and Scientists (WDTS) under the Science Undergraduate Laboratory Internships program for supporting this research. Further work of A.S.D. was supported by the US Department of Energy, Office of Science, Basic Energy Sciences, Materials Sciences and Engineering Division under Award No. DE-SC0018660. Y.C. and J.A.M.P. were supported by the Scientific User Facilities Division, Office of Basic Energy Sciences, U.S. Department of Energy, under Contract No. DE-AC0500OR22725 with UT Battelle, LLC.
\end{acknowledgments}

\end{document}